\documentclass[sigconf]{acmart}
\usepackage{comment}

\AtBeginDocument{%
  \providecommand\BibTeX{{%
    \normalfont B\kern-0.5em{\scshape i\kern-0.25em b}\kern-0.8em\TeX}}}

\setcopyright{none}
\acmDOI{}
\acmISBN{}
\acmYear{2024}

\acmConference[LIMITS '24]{Tenth Workshop on Computing within Limits}{June
18--19, 2024}{Online}
\begin{document}

\title{Computing, Complexity and Degrowth : Systemic Considerations for Digital De-escalation}

\author{Valentin Girard}

\affiliation{%
  \institution{Laboratoire d'Informatique de Grenoble}
  \streetaddress{Bâtiment IMAG, 700 Av. Centrale}
  \city{Saint Martin d'Hères}
  \country{France}
  \postcode{38 401}
}
\email{valentin.girard2@univ-grenoble-alpes.fr}

\author{Maud Rio}
\affiliation{%
  \institution{Laboratoire G-SCOP}
  \streetaddress{46 Av. Félix Viallet}
  \city{Grenoble}
  \country{France}
  \postcode{38 000}}
\email{maud.rio@g-scop.eu}

\author{Romain Couillet}
\affiliation{%
  \institution{Laboratoire d'Informatique de Grenoble}
  \streetaddress{Bâtiment IMAG, 700 Av. Centrale}
  \city{Saint Martin d'Hères}
  \country{France}
  \postcode{38 401}
}
\email{romain.couillet@univ-grenoble-alpes.fr}


\begin{abstract}
Research on digital degrowth predominantly critiques digital expansion or presents alternative digital practices. Yet, analyzing the link between digital technologies and complexity is crucial to overcome systemic obstacles hindering digital de-escalation. This article presents the different types of links between complexity and computing observed in the literature: the infrastructural complexity inherent in digital technologies, the socio-political complexity induced by them, and finally, the ontological complexity (individual's ways of relating to their environment) hindered by digitization. The paper explores these links to identify ways to reduce infrastructural and socio-political complexities, and to move away from the reductionist paradigm, in order to support digital degrowth. Its development shows that complexity induces ratchet effects (i.e. irreversibilities in the development of a technique in a society), rendering degrowth efforts difficult to handle by individuals. Therefore, strategies to overcome these barriers are proposed, suggesting that bottom-up simplification approaches stand a greater chance of making alternatives emerge from different stakeholders (including users). This digital shift assumes the development of methods and technical tools that enable individuals to disengage from their attachments to digital habits and infrastructure, opening a substantial field of study.
\end{abstract}

\keywords{Digital de-escalation, Degrowth, Complexity, Infrastructure, Ratchet effect, Simplification} 


\maketitle
\section{Introduction}

The concept of degrowth is increasingly documented in the academic field. Originating in the 1970s, employing arguments of necessity with an aim to influence the economic sphere (see for instance \cite{meadows_1972,georgescu_1971}), it re-emerged predominately in France during the early 2000s (see for instance  \cite{latouche_2006, rabhi_2010}), proposing a new societal model based on a paradigm shift in individual's relationship with the world, more focused on the synergy between the environment, society, and people \cite{lievens_2022}. Around the 2020s, particularly since the COVID crisis, degrowth has experienced a significant rise. Parrique offered a new definition for the degrowth concept: "a reduction in consumption and production in order to lighten ecological footprint, democratically planned with a focus on social justice and well-being" \cite{parrique_2022}. In this position the degrowth movement transcends a simple opposition to growth: embodying a political agenda for an alternative societal organization, associated to another life philosophy. Degrowth movements are definitely not centred on reducing GDP, but aspire to the desire of moving beyond the dependence on growth, prioritizing alternative indicators such as happiness, equality, access to public services, etc. This international movement is gaining strength all over the world\footnote{This article predominantly source French references, aligned with international visions of the theoretical and strategic development of the degrowth concepts.} (see for instance \cite{jackson_2011,kallis_2020,hickel_2020}). 

In practice, digital degrowth (i.e. specification of degrowth for the digital sector) is challenging. The staggering statistics regarding the speed of its development and impacts\footnote{+8\% annual increase in greenhouse gas emissions, +9\% annual increase in energy consumption, +40\% annual increase in data storage, +25\% increase in data flow \cite{lean_2018}.} make the prospect of a reduction in the production of digital hardware unlikely. This obstacle is further fortified by the perception that digital solutions are often deemed indispensable for societal organization, or that "the global systemic effects of the current digital transition [...] are often considered positive a priori" \cite{lean_2018}.

By contrast, arguments related to the sustainability of the digital infrastructure \cite{monnin_2020}, scarcity of metallic resources \cite{bihouix_2021}, ecological impact \cite{lean_2018}, social impact \cite{chabanne_2023}, societal stability \cite{gisondon_2020}, and ethics \cite{leonarduzzi_2021} increasingly call for digital sobriety \cite{bordage_2019,lean_2018} or digital de-escalation Digital de-escalation, defined as the counteraction to digital escalation, aims at slowing down, stabilizing, or even decreasing the digitization of our society. This requires not only reducing the production of hardware but also fostering a collective detachment from these technologies. \cite{couillet_2023}. Additionally, the field of research on the link between degrowth and digital de-escalation is gaining depth, proposing studies on the synergies between these two concepts \cite{espana_2023,kerschner_2018,sharma_2023}  or highlighting initiatives and experiments of digital solutions that deviate from the growth paradigm \cite{abbing_2021,sutherland_2022}. The position of analyzing socio-environmental impacts and limits of growth in the digital sector, and the elaboration of alternatives to unrestrained growth are widely supported and documented by the \textit{Computing Within Limits} community \cite{nardi_2018}.

While such initiatives are crucial for advancing the field of research, if not contextualized within a systemic framework, they will be vulnerable to strategic obstacles that could hinder their implementation on a larger scale. Critiques of the Sustainable Human-Computer Interaction (SHCI) research field have often emphasized the need to broaden research to a more systemic approach and to explore the concept of degrowth \cite{bremer_2022}. It appears important, therefore, to revisit the systemic dynamics that oppose the widespread adoption of digital degrowth and emphasize the significance of understanding these dynamics to avoid being powerless in the face of macro-societal phenomena. Anchoring this reflection in the work of the \textit{Computing Within Limits} community, this paper intents to bring new systemic considerations into the community focus. This will enable the identification of strategic levers for a paradigm shift in the digital realm. Thus, studying the link between digital technology, complexity, and degrowth is envisaged in this paper as a pertinent manner to understand one facet of systemic effects and overcome them.

The concept of complexity is broad and can encompass several definitions. This term generally refers to the number and diversity of entities and the relationships between these entities within a system. However, depending on the system, the entities, and the relationships considered, the nature of complexity is entirely different. For example, the complexity of an ecosystem, a technology, the human brain, a society, etc. are addressed in their respective field of study. 

In the digital related fields, the term \textit{complexity} is used in polysemous ways, particularly in literature related to digital degrowth: sometimes to denounce the role of the digital sector in societal acceleration, the verticalization of powers, or to highlight its complexity in manufacturing and network operation. On the other hand, digital technologies are sometimes described as interfering with the complexity of non-virtual human relationships, with the relationship of humans to nature, or even symbolizing a form of homogenization and simplification of the world, thus opposed to the complexity of ecosystems for instance. The primary aim of this article is therefore to redefine which forms of complexity are associated with digital technologies, in order to (a) understand the different obstacles these forms of complexity pose to digital degrowth, and (b) identify which aspects of complexity can be leveraged to trigger and organize degrowth.

In this view, this paper examines three prisms of vision of complexity and explore the way this visions are linked to the concepts of degrowth and digital de-escalation. The research methodology applied in this paper is outlined in Section 2 and has been elaborated to address the following research question: “how digital de-escalation can be addressed considering the various forms of complexity associated with the computing fields ?”. 
Section 3 explores the connections between digital technologies and complexity in the literature. Section 4 discusses the implications of these understanding for degrowth, and Section 5, addresses three barriers to simplification through the lens of the "ratchet effect" concept. Finally, strategies for overcoming these ratchet effects are proposed in Section 6.

\section{Research Methodology}
\subsection{Context of the Research Study}
This study is conducted in the context of a PhD research project bridging the fields of design sciences (in a broad sense: including a socio-technical system engineering perspective, and related methodologies) and the critique of digital escalation (from a material, computational, including more holistic perspectives, and related experimentation with actors). This study is thus situated at the intersection of these disciplinary fields. The methodology used for this thesis is the Design Research Methodology \cite{DRM}, and this article is positioned in a phase of Research Clarification, based on a deep literature analysis.

\subsection{Research Fields Exploration} 
The first phase of this research explored the results of a project leaded by five students at ENSE3 School in Grenoble (France), entitled "Low-tech computing" (unpublished). In this prior work, the paradoxical relationship between the two concepts of "low-tech" and "computing" was analyzed to provide a first state-of-the-art overview on the topic, and to propose, as a demonstrator of an alternative way of supporting similar service, a prototype of a web server hosted on a salvaged smartphone, powered by photovoltaic energy. This exploration generated a bibliography that served as the foundation for this study. The bibliography was further expanded, based on references cited by key researchers in the field, such as Gauthier Roussilhe, Alexandre Monnin, Thimothée Parrique, Laurence Allard, José Halloy, Ivan Illich, and others. 

In total, 16 books, 4 reports, 7 web contents (videos and podcasts), and 13 articles were studied during this phase. This exploration identified various sub-domains that subsequently served as the prism of analysis:
\begin{enumerate}
    \item Critique of the current digital paradigm
    \item Complexity
    \item Degrowth
    \item Alternatives to the current digital paradigm
    \item Ecological redirection (related to the study of \textit{attachments} and \textit{renunciation}).
\end{enumerate}
It is noteworthy that themes related to degrowth and ecological redirection are predominantly explored in France.

This exploratory phase identified the significance of the challenges of digital de-escalation. This phase has notably highlighted the entanglement of the themes of digital de-escalation and alternatives proposed to the current digital landscape with the concept of degrowth. However, the degrowth paradigm must be approached with a systemic analysis to detach from the growth paradigm. Thus, the question of complexity serves as an interesting analytical lens due to its systemic approach and strong connection with digitization. This observation led to the completion of a analysis of the literature on this topic.

\subsection{Literature Analysis}

To avoid missing important articles in this domain, a literature analysis was conducted. The previously mentioned sub-domains (1), (2), and (3) served as the basis for constructing a keyword search. The following equation was used as the search query in the databases \textit{Google Scholar}, \textit{ACM}, \textit{Bib CNRS}, and \textit{Web of Science}:

(1): (Digital OR numérique OR comput*) AND (degrowth OR "post-growth" OR "post-croissance" OR décroissance) AND (complex OR complexité OR complexity)

As a selection criterion, searches yielding more than 200 results were refined. This was done in the Google Scholar database, where only review articles were selected, and theme (4) was added by completing the search equation:

(2): (Digital OR numérique OR comput*) AND (degrowth OR "post-growth" OR "post-croissance" OR décroissance) AND (complex OR complexité OR complexity) AND "low-tech"

Following this collection of articles, three exclusion criteria were applied:
\begin{itemize}
    \item Articles not addressing computing as a central theme
    \item Articles rooted in the current digital paradigm and/or a growth paradigm
    \item Articles not treating the digital question holistically.
\end{itemize}

After this selection process, 15 articles were retained. Table \ref{table-literature} summarizes the article collection phase.

\begin{table}[h!]
  \caption{Number of articles for each stage of the article collection.}
  \label{table-literature}
  \begin{tabular}{lccc}
    \toprule
   Data base&Equation (1)&Equation (2)&After exclusion\\
    \midrule
    Google scholar&31 700&\textbf{34}&3\\
    ACM&\textbf{42}&&9\\
    BibCNRS&\textbf{172}&&2\\
    Web of Science&\textbf{12}&&2\\
  \bottomrule
  After eliminating&&&15\\
  duplicates&&&
\end{tabular}
\end{table}

\section{Complexity and computing in literature}

The digital sector has a strong connection with complexity. This relationship has the particularity of being bi-directional: complexity supports the unrestrained development of the digitalization, and this development, in return, promotes the complexification of Western societies. Indeed, the complexity of a typical french-type society is linked to globalization, and the intensity of its tight-flows economy based on growth. The economy is based on a worldwide system promoting intensive growth of flows extracted, used and discarded. This complexity today supports the demand for an intensive raw materials usage, used for the production of digital materiality.

Conversely, digital technologies enable the support and development of this global complexity by fostering the exchange of information and thus optimizing the organization of  societies, which tends to grow in complexity through rebound effects\footnote{vidence and arguments against green growth as a sole strategy for sustainability were clearly established in 2018, see: https://eeorg/library/decoupling-debunked/}.

On the other hand, some authors view digital technology more as an opposition to the complexity of human's relationship with with each other and with other living beings. Digital technology would then be considered as the anti-complexity, aiming to homogenize the interdependent relationships between humans and with living beings, thus advocating for ontological reductionism.

These kinds of connection between computing and complexity appears in the literature and will be developed in the following subsections. As complexity is discussed here in a generic sense, covering various definitions, it needs to be specified. The co-occuring forms of complexity are not independent of each other, and their dynamics share common foundations. The following subsections study separately the infrastructural, socio-political and ontological complexities inherent to the digital sector.

\subsection{Understanding Infrastructural Complexity}\label{infrastructural_complexity}
The infrastructural complexity refers to the complexity of devices that constitute digital technologies, the complexity inherent in the value chain that generates them, and the complexity of their interconnection. 

These devices are based on the most globally sophisticated technologies worldwide. Their material composition stands on up to 50 different chemical elements \cite{pitron_2021}. A vast number of components and sets of components are interacting with each other. In addition, the manufacturing processes required to fabricate these devices are highly complex. New digital devices are designed with billions of transistors, each only a few nanometers in size. These transistors, composing electronic chips, are coupled with many other components, each smaller and more efficient than previous versions (capacitors, inductors, batteries, LCD screens, touchscreens, cameras, microphones, motors, etc.) exchanging electrical signals to form a coherent system capable of receiving, processing, and transmitting information abundantly and rapidly.

Infrastructural complexity also stems from industrial value chains that enable the production and support of this downstream infrastructure. These downstream terminals and networks exist thanks to the most complex supply and distribution chain globally.
For example, the life cycle of an electronic chip for smartphones begins in mines (upstream flows). For most of the components, the majority of these mines are located in global South countries. Mines are based on heavy mechanical and chemical processing, generating severe environmental, health and social consequences for the inhabitants of these countries \cite{izoard_2024,boluda_2021}. China has almost the monopoly on the rare earth market \cite{lean_2018} and produces the majority of the chips used in the world. Smartphone chips have transistors etched at less than 10 nanometers. Only one photolithography machine (based on Extreme Ultra Violet light) -- costing several hundred million dollars, making it one of the most expensive machines in the world -- developed by the company ASML in the Netherlands is capable of etching such chips onto silicon disks (wafers) \cite{miller_2022}. The Taiwan Semiconductor Manufacturing Company (TSMC) factory operating in Taiwan leads the production of these technology based chips. So-called "white rooms" are used to manufacture the chips, generative intensive environmental and societal impacts : a daily water consumption of 15,000 tons (for washing wafers numerous times with ultra-purified water) which creates conflicts of use with residents and farmers \cite{reporterre}, as well as very high electricity consumption (largely carbon-intensive in Taiwan), and chemical pollution of downstream rivers. Smartphones assembled in China, for example, in the Foxconn factory, involve working conditions that are akin to modern slavery \cite{izoard_2020}. 

The electronic devices design phase is usually operated in Western countries, whereas the manufacture is rather in Asia. The devices are transported to consumers to be used for only a few years before becoming non-functional or simply outdated. The majority of electronic devices and small components, are not designed to recycle their embedded materials in a cost-effective manner. The 54 million tons of electronic waste generated in 2020\footnote{https://ewastemonitor.info/gem-2020/} (equivalent to the weight of 7,400 Eiffel Towers) will be recycled for only 17\% of them \cite{forti_2020}. The remaining 83\% will end up in landfills, mostly in Africa and Asia, often through illegal trafficking. 

On a larger scale, infrastructural complexity is manifested by the hyper-complex networking of these terminals. This networking is enabled by electrical cables, optic fiber cables, antennas, and transformers. Data centers allow for the storage of increasingly externalized information from users' devices, further maximizing data exchanges on the network. This infrastructure is increasingly globalized and hierarchized, with distribution points and high-level connection lines, redistributing information locally through lower network layers. This network, characterized as a Large Technological System (LTS) by Lopez, is defined by elements such as "large scale; a specific development mechanism favoring their growth; the consumption of significant amounts of fossil or fissile energy; the production of high CO2 emissions or ultimate wastes; overall complexity and sub-ensembles operating in networks, often opaque and centrally managed by experts; real-time control and regulation of flows" \cite{lopez_2022}.
    
\subsection{Analyzing Socio-political Complexity}

This infrastructural complexity maintains a strong connection with socio-political complexity, and the digital sector, as an organizational tool at all levels of society, plays a significant role in the socio-political complexification of our society. Socio-political complexity has been extensively studied by Joseph Tainter in his work \textit{The Collapse of Complex Societies} \cite{tainter_1988}, referring to "notions such as the size of a society, the number of its components and their characteristics, the variety of specialized social roles it encompasses, the number of distinct social personalities present, and the variety of mechanisms to organize them into a coherent and functional whole" \cite{tainter_1988}. In a recent article, Tainter also mentions components such as technical capabilities, hierarchy, and the quantity of information produced and transmitted \cite{tainter_2006}. 

This digital realm stands on a set of technologies and infrastructures that enabled today's level of socio-political complexity. While infrastructural complexity allows the digitization of the world, the digitization of the world contributes to the development of socio-political complexity. Complexity is still not inherently negative and serves a purpose for societies: it eases them to address various issues and, most importantly, ensures increasing living comfort. However, such complexification always comes at a cost: through an increase in energy and raw material consumption. For digital technologies, this cost is immense today, requiring a vast amount of mining raw materials and fossil energy to operate.

Thus, digital development is the \textit{sine qua non} condition of the new level of complexity reached at the end of the 20th century. The upgraded characteristics of socio-political complexity mentioned by Tainter can be summarized as follow:

\begin{enumerate}
    \item\textbf{The size of a society} : the boundaries of the globalized society are no longer limited to habitable continents. Due to digital technologies usage, boundaries are extended to previously uninhabitable places (for example, the South Pole or space) and even virtual worlds”. The metaverse created by various games, applications, and social networks expands the world into a fictional space represented by various semiotic means (videos, texts, audios, social networks, communities of internet users, etc.).
    \item\textbf{The number of its components and their characteristics} : progress in the health domain, facilitated by digital technologies, has  prolonged lifespans and reduced premature deaths, contributing to the increase in the global population. The digital itself consists of a very large number of autonomous machines communicating information, becoming full-fledged entities in the analysis of complexity. The 20 billion connected devices communicating over the internet and the Internet of Things \cite{pitron_2021} (in 2018, likely more at the time of reading) represent new entities increasing the number of "agents". Characteristics are no longer solely human but also algorithmic, demanding new forms of interactions between machines or "human-machine" interactions.
    \item\textbf{The variety of specialized social roles it encompasses, the number of distinct social personalities present, and their hierarchy} : the digitization presupposes an increasingly vertical, hierarchical, and dominant system, implying an extension of the categorization of individuals on this dominance scale. The manufacturing of our digital devices accentuates inequalities by creating a real system of domination on a global scale. As said by Monnin : "Technology [...] crystallizes unequal flows of energy, matter, and work. Its development does not cause inequalities but presupposes them" \cite{monnin_2022}. Moreover, the rise of comfort and available time due to problem-solving facilitated by new digital technologies (heating, construction, agriculture, etc.) has led to a shift of professions towards tertiary activities, creating new roles in societal organization.
    \item\textbf{The variety of mechanisms to organize them into a coherent and functional whole }: the societal organization facilitated by digitization is a key point in the analysis of socio-political complexity induced by digitization. Digitization has not only allowed the complexification of organizational mechanisms in government and intergovernmental organizations (extension and complexification of administrative procedures through digital means, instant communication with populations, intergovernmental communication enabled by this infrastructure), but also and especially, enabled the organization of various smaller social groups: a group of friends, a family, an association, a company, a municipality, etc. These new forms of organization involve a different relationship with time, space, and relationships. We pay online, get deliveries, make instant appointments, change plans at the last minute, inform of delays through a voice message, reserve, share, sell second-hand items, stay informed, video call, engage in long-distance intimate activities through connected toys, play online games. Small-scale social organization becomes more and more connected. Non-digital individuals have no say in a society where everything happens on the internet \cite{france_inter}.
    \item\textbf{Technical capabilities} : unmatched in the digital age. See section \ref{infrastructural_complexity}.
    \item\textbf{The quantity of information produced and transmitted}: which is the basic purpose of digital technologies, designated as New Information and Communication Technologies (NICT). Digital data grows exponentially, with staggering growth rates: + 25\% per year for data flow (resulting in a threefold increase every 5 years), +40\% for data center storage \cite{lean_2018}. 
\end{enumerate}

\subsection{Addressing Ontological Complexity}
The infrastructural and socio-political complexity supporting and induced by the digitization of the globalized economy is often criticized for its growth. Digital technologies and complexity are then seen as being caught in a vicious cycle oriented towards infinite growth. However, other authors see the digitization of the world as a reduction of complexity, referring here to human's relationship and conscientization of the living world, human's social relations, etc. This corresponds to \textit{ontological complexity}.

This form of complexity corresponds to a way of perceiving the world that is opposed to the reductionist one. The latter often refers to the combination of three principles: exclusion, reduction, and disjunction: "all these traits have in common a paradigm of exclusion, which simply excludes from scientificity, and thereby from 'true' reality, all the ingredients of the complexity of reality (the subject, existence, disorder, randomness, qualities, solidarities, autonomies, etc). The paradigm of exclusion is associated with a principle of reduction which enjoins to disintegrate global entities and their complex organizations in favor of the elementary units that constitute them. [...] To this was added an internal paradigm of disjunction, which isolated sciences from each other and, within these sciences (physics, biology, human sciences), disciplines from each other, abstractly and arbitrarily cutting their object from the solidarity fabric of reality." \cite{morin_1991}. 

Different authors argue that digital technologies contribute to this reductionist paradigm by depriving individuals of access to the complexity of the living world and non-virtual human interactions, imposing a reductionist vision of this complexity. Overexposure to an addictive virtual universe leads some individuals to develop a \textit{nature-deficit disorder} \cite{louv_2008}, a psychological disorder that makes natural environments appear hostile to those affected. This disorder arises from the loss of contact with natural environments, the breaking of ties with the living world, and the loss of rooting in ecosystems adapted to humans. Couillet et al insist on the fact that our "\textit{cosmology}" is disrupted by virtual spaces \cite{chabanne_2023}. Barrau also highlights this phenomenon, pointing to a purely scientistic ontology of Western societies that confine themselves to mechanisms of societal digitization and techno-solutionism, excluding other more sensitive and complex visions of reality \cite{barrau_2022}.

\section{Implication for degrowth}
The link between computing and degrowth thus appears as an opposition depending on the perspective adopted to describe it. This paper investigates the polysemous term "complexity" and its connections with computing to grasp its implications on societal dynamics. This section analyses the evolution of different forms of complexity in a society shifting towards degrowth.

Digital degrowth cannot be reduced to a decrease in the production and consumption of digital goods and services. As degrowth in its general sense, the movement involves a genuine ecological redirection program for these sectors, reorienting them towards objectives of well-being and reducing inequalities, within a democratic framework, while ensuring their compatibility with planetary boundaries and the preservation of ecosystems. This movement implies arbitration policies definition, reallocation of infrastructure, new modes of governance, etc. Thus, questioning the implication of complexity in a broader digital degrowth program is necessary to establish a theoretical framework that subsequently allows for the proposal of action levers consistent with degrowth.

\subsection{Simplification as an alternative to collapse}

In his substantial work, Tainter provides an analysis of the collapse of complex societies in the past, attempting to offer a general explanation based on complexity. By collapse, it is essential to understand "a rapid and decisive loss of an established level of socio-political complexity" \cite{tainter_1988}. 

According to him, societies that have collapsed follow a similar scenario. The increase in socio-political complexity enables societies to solve the problems they encounter (climate constraints, invasions, shortages, etc.) and to gain in comfort : "overall, we are better off because we have become complex. Complexity is obviously very useful for solving problems" \cite{tainter_2006}. Lowering complexity is often not possible, either due to irreversibilities in development or the elite's desire to remain seen as legitimate. 
Complexity continues to increase, whereas marginal returns (= benefit produced / associated cost) are decreasing in all areas (resources, energy, conquests, education, etc.). However, complexity comes at a cost: energy. Whether from an external source (renewable, fossil, colonized peoples) or the labor force of the society in question (paid or not \cite{mies_1986}), more complexity implies more energy expenditure: "The cost of complexity is the energy, labor, money, or time required to create, maintain, and replace systems that involve more and more parts, specialists, behavior regulators, and information" \cite{tainter_2006}. Individuals must invest more resources in supporting complexity. Once budget reserves used to maintain the status quo are exhausted and that complexification efforts no longer produce significant effects because of the fall in marginal returns, the system can no longer withstand shocks. This leads to a rapid collapse to a lower level of complexity which allows the reappearance of increasing marginal returns. This has political consequences, but also social (living conditions) and cultural ones.

\begin{figure}[ht]
    \centering
    \includegraphics[width=\linewidth]{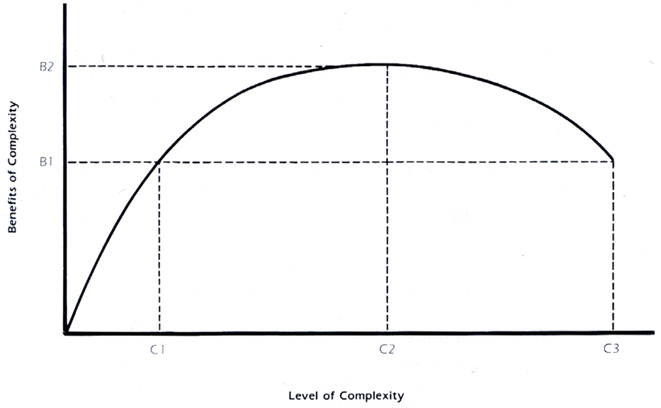}
    \caption{Evolution of the benefits of complexity for a complex society (extracted from \cite{tainter_1988})}
    \label{complexity_curve}
\end{figure}

In the case of digital technologies, fossil fuels (especially oil) and mining extraction mainly support the high level of complexity brought to the Western society. 
The marginal return of mining extraction is however diminishing. For example, the average concentration of copper mines decreased from 1.8 kg/ton at the beginning of the 20th century to 0.8 kg/ton at the beginning of the 21st century \cite{bihouix_2010}. This decrease in return induces greater energy consumption. The production of energy is also subject to diminishing returns, especially for oil. Although it is challenging to estimate a date for the peak or end of mining extraction \cite{geldron_2017}, it is reasonable to think that this decrease makes society highly vulnerable. The decline in these marginal returns does not, on its own, cause a collapse, but it creates vulnerability to shocks that can no longer be overcome by increasing complexity since the benefits decreases (see figure \ref{complexity_curve}). Thus, the decrease in marginal returns of digital infrastructure sets the conditions for a collapse triggered by shocks that could be related to the availability of water, the health, food, and political situation of the Southern countries on which the infrastructure depends, the disruption of a critical supply chain, etc. Tomlinson imagined an example of a collapse scenario triggered by consequences of climate change : "the indirect global effects of climate change could be so dramatic (e.g., wars, famines) that significant reductions in sociotechnical complexity may have led to a great degree of deglobalization" \cite{tomlinson_2017}.

This analysis of collapses, however, must be nuanced. Indeed, Tainter's theory is based on societies much less complex and vast than Western society, which extends over the majority of the globe and is based on a globalized economy. The interdependencies complexity between countries is strong on diplomatic and commercial levels, which creates structural differences from the societies studied by Tainter. Furthermore, modern Western societies have an economic model largely based on innovation, which according to Tainter allows for a modification of the curve of the benefits of growth, thus delaying the decline of marginal returns \cite{tainter_1988}. Lastly, the analysis conducted in this paper has been solely based on the digital sector. An additional analyze of the dynamics of complexification and gains brought in numerous sectors is necessary to formulate reliable conclusions regarding vulnerability to collapse\footnote{However, it is important not to exclude the hypothesis of collapse propagation due to cascade effects and irreversibility: a digital collapse could greatly affect socio-political organization with runaway phenomena, especially in highly digitized countries.}. These forms of digital infrastructure collapse are addressed in the literature, notably by Tomlinson who introduced the concept of \textit{peak ICT}\footnote{a peak in the production of ICT, followed by a decline} \cite{tomlinson_2017} and \textit{Collapse Informatics}\footnote{a type of digital tools that adapt to collapse and enable to face it} \cite{tomlinson_2013}, and by Jang who describes the potential fate of digital infrastructure following a forced cessation of hardware renewal \cite{jang_2017}.

Nevertheless, while the increase in socio-political complexity jeopardizes the stability of today's worldwide economy-based society, collapse may not be the only outcome. Tainter proposes in a more recent article \cite{tainter_2006} to consider three different scenarios for institutions to address long-term problem solving: "these scenarios are collapse, resilience and recovery through simplification, and sustainable problem resolution based on increasing complexity subsidized by new resources." In the case of the digital realm, the possibility of increasing complexity based on new resources is likely excluded because six of the nine planetary boundaries defined by researchers from the Stockholm Resilience Center \cite{rockstrom_2009} are being crossed in 2024 \cite{richardson_2023}. Resilience through voluntary digital simplification could therefore remain an option to avoid a collapsing scenario. It requires a sacrifice (in a salutary sense) of society's durability since it involves mutating to adapt to the problems it faces rather than overcoming them. 
However, this loss of durability prevents a society from collapsing. Referring to Tainter's illustrative case: "the Byzantine example was a resilient response because it involved abandoning long-established traditions of government and society. In this case, costs are reduced, and the productive system is improved. This is a strategy that, in the Byzantine case, allowed for budgetary recovery and, ultimately, expansion" \cite{tainter_2006}.

A strong link between degrowth and decomplexification is clearly identified in this example: since an increase in complexity requiring an increase in the energy and resource consumption, is no longer viable, the "reduction in production and consumption to lighten our ecological footprint" \cite{parrique_2022} associated with the concept of degrowth shall be combined with \textbf{a reduction in socio-political complexity to lighten societies' energy and resource consumption and increase their resilience}.

A degrowth analysis through simplification is indeed not new, as the movement already advocates values in favor of simplification. It often stands for a return to plural societies based on localism \cite{kostakis_2023}, in order to reduce interdependence and hierarchy.
The relationship between degrowth and technology, although diverse in response to the question of technique \cite{kerschner_2018}, strongly encourages initiatives related to low-tech, the simplification of technology usage, and the logistic chains that ensure their production. The degrowth movement also advocates for a reduction in the energy intensity and in the material consumption, which, in itself, contributes to complexity. The imaginary surrounding degrowth often revolve around the connection to one's immediate environment, a calming of life intensity, and convey a vision of simplification as having a very positive impact on human health \cite{parrique_2022}. These visions evoke tranquility, serenity, a return to emotions, and reflection, portraying simplification as a joyful process.

Therefore, the notion of collapse does not necessarily refers to apocalyptic scenarios\footnote{In the case of our society, one could even imagine that this reorganization could curtail the death of thousands of billions of animals each year. Nevertheless, a decrease in human population is not to be excluded, whether through a decline in birth rates or the death of numerous people.}, but entails at minimum a radical shift in the lifestyles of the individuals composing the society in question, which entails a socio-political structural reorganization based on less complexity \cite{tomlinson_2013}.


\subsection{Shifting paradigm for degrowth}

Although degrowth must be supported by a simplification of the infrastructural and socio-political complexity induced by digital technologies, it should also be accompanied by a new ontological shift, oriented towards the complexification of human's relationship with the world.

Lievens considers that the new form of the degrowth movement that reemerged in the 2000s fundamentally differs from the movement of the 1970s by placing itself within a \textit{metaparadigm of complexity} "which would allow to handle this complexity, something impossible for the simplification paradigm." \cite{lievens_2022}. Indeed, the emergence of the degrowth movement claimed the end of economic growth to decrease impacts on nature, thus remaining in a logic of human/nature distinction \cite{descola_2005}, without questioning today's (mainly Western) society ontologically. Changing trajectory while remaining within a \textit{metaparadigm of simplification} comes down to making adjustments in a well-defined system, without questioning the system itself. It "takes place within a system (whatever it may be) and operates by modifying certain variables." \cite{lievens_2022}. For Lievens, the degrowth movement of the 2000s, which he calls \textit{neo-degrowth}, differs in that it seeks to break free from the \textit{metaparadigm of simplification} and advocate for a \textit{type 2 change}: "this change no longer takes place within the given framework or system, but concerns the framework itself, by redefining it in its structures." \cite{lievens_2022}. According to him, the \textit{neo-degrowth} movement, which relies on three paradigms in interaction (ecologist\footnote{"The ecologist paradigm is the one that questions how humans interact with their natural environment."}, societalist\footnote{"The societalist paradigm is the one that questions how humans form society."}, and subjectivist\footnote{"the subjectivist paradigm supports the other two by putting the question of the meaning of human existence in the background."}), effectively inserts it into a \textit{metaparadigm of complexity}, requiring a \textit{type 2 change} in today's main axiology.

In this regard, \textit{neo-degrowth} differs from the 1970s degrowth movement as it mobilizes a different leverage point, based on societal paradigm shift rather than system direction\footnote{This paradigm shift has already been documented in the \textit{LIMITS} community by Comber \& Eriksson \cite{comber_2023}.}. According to Donella Meadows, there are 12 leverage points, which represent ways to influence the behavior of a complex system (such as our society). They can be hierarchized from the least transformative (12) to the most transformative (1) \cite{meadows_1999}. Thus, the 1970s degrowth movement can be associated with leverage point number 3: a willingness to challenge the intrinsic goal of our society\footnote{which does not refer to the average personal goals of the individuals comprising it but rather the goal of the system itself, which may differ.}, namely growth. \textit{Neo-degrowth}, on the other hand, proposes to go one step further, to leverage point number 2, thereby focusing on societal (meta)paradigm shift, which is "the source of the system".

To recap, degrowth maintains a plural relationship with the notion of complexity: this movement advocates for an infrastructural and socio-political simplification of our Western societies, while claiming a \textit{type 2 change} on an ontological level to reach a \textit{metaparadigm of complexity}. In the following sections, the notion of paradigm shift is let behind to focus on barriers that are opposed to a simplification of infrastructural and socio-political complexity.

\section{Understanding the Ratchet Effects}

Since a decrease in socio-political and infrastructural complexity is desirable, this research paper investigates the mechanisms that oppose this simplification. One of such mechanism (in line with the systemic analysis addressed in this research) could be found in the notion of \textit{ratchet effect}. This refers to a phenomenon of irreversibility, often used in economics: "if wages increase due to inflation, they will not decrease if prices fall"\footnote{https://www.alternatives-economiques.fr/dictionnaire/definition/97138}; or in sociology : "consumption that has been achieved is difficult to reduce due to habits and commitments that have been made" \cite{duesenberry_1962}. The term is not frequently used in the study of complex Western societies. For the digital case, this could be interpreted as an inability to revert once the infrastructure has been developed to a certain stage, as the cost of dismantling is (perceived as) much too high compared to the benefits it can bring. It is worth noting that these phenomena of irreversibility in complexification are described in the literature, but not under a generic term. A literature search points to different terms such as "path dependence", "lock-in effect", and so on.

In the literature studied for this research, three types ratchet effects related to complexity seems to emerge. As with complexity, these effects may not be considered interdependent, but they have in common foundations and overlaps in their field of action. They result from complex global phenomena that carry them in the same dynamic.

\subsection{Social Ratchet Effect}
The first directly concerns individuals. Habits and the comfort gained from the complexity of our society inhibit any individual will to revert. This phenomenon is particularly due to \textit{attachments}, a term that is understood as "what we care about and what holds us", especially strong in the context of digitization \cite{podcast}. This effect works both ways. 

On the one hand, individuals are attached to the digital technologies offered to them: "the smartphone becomes the unique vector of an \textit{affective condensation} without equivalent. How can we be surprised that we feel attached to it as the apple of our eye, since the news of the world, as well as the most important communications related to our job, our bank account, our idols, our friends, our parents, our children, our lovers, all arrive in a continuous flow through it ?" \cite{citton_2022}

In the opposite direction, digital technologies create a dependence to itself. 
This dependence is to be understood by the master-slave relationship that has reversed for the case of digital and humans. Illich explains this by the notion of radical monopoly: "a type of domination by a product rather than that of a brand. In such a case, an industrial production process exerts exclusive control over the satisfaction of a pressing need, excluding any recourse, for this purpose, to non-industrial activities" \cite{illich_1973}. The digital realm is a hyper-radical monopoly; it is the only way -- at least in our Western society -- interact with the rest of the social organization, thus being included as an individual in society. Without digital tools, it is almost impossible to maintain a contractual relationship with the state, and therefore to have a social contract. Socialize get challenging if the entire society maintains a relationship with time and a lifestyle different from mine due to a technology. The digital realm has the radical monopoly of social organization: it manages schedules, allows communication with others, remotely and at any time, models action plans, engraves memories, alerts on an event or on the news. And as Illich points out, the radical monopoly threatens the autonomy of the individuals and transforms them into a tool slave, making us incapable today of doing without it. This dependency creates irreversibility in the development of the digital sphere.

\subsection{Technological Ratchet Effect}
Another type of ratchet effect is related to the infrastructure itself. Since digital technologies act as a network, they are more susceptible to "path dependence," which means a situation where "a minor or fleeting advantage or a seemingly inconsequential lead for some technology, product, or standard can have important and irreversible influences on the ultimate market allocation of resources, even in a world characterized by voluntary decisions and individually maximizing behavior" \cite{liebowitz_1995}. This phenomenon, which leads to significant societal lock-ins, was explained by Arthur \cite{arthur_1994} through the theory of increasing returns and network effects. Indeed, a network-based technology provides a benefit to its users that grows with the number of users. Thus, a phenomenon of positive feedback is created by increasing marginal returns: a slight market lead of a technology will increase its benefit to users. Consequently, the increasing returns from early gains tilt the competition towards the technology that accumulated advantages early enough to establish itself as a (quasi-)monopoly in the market ("the winner takes all").

Recently, Lopez has developed one step further this idea. In this view, Large Technological Systems are guided by an expansive force \cite{lopez_2022}, the power of which inhibits any reversal. The ratchet effect related to the digital infrastructure is, therefore, strongly linked to the network it forms. Ultimately, it is the actors forming this network that work towards its expansion and reject any unpredictable element against the norms they have set. This desire for expansion and complexification is, consistently to Arthur's theory, guided by an efficiency objective: "the mega-machine expands like an infernal infrastructure whose locks multiply and shift. The disabling efficiency of this complexity delegitimizes any will for deconstruction and structural transformation" \cite{lopez_2022}.

\subsection{Political Ratchet Effect}

Finally, an explanation for the ratchet effect related to complexity could be the power play established by current highly hierarchical (e.g. Western) societies, allowing an elite to concentrate power. Tainter explained in the theory of the collapse of complex societies that for an elite, rolling back political measures of comfort granted to citizens, for reasons of legitimacy, is highly challenging. Similarly, this quest for legitimacy prevents elites from rolling back technological advancements when the cost of dismantling cannot justify short-term benefits. If decisions are made by elites, they must be made in the interest of those relinquishing decision-making power. Otherwise, subordinates will no longer perceive the legitimacy of this elite going against their will, and it will be overthrown if the repression it exerts on the population is insufficient. In that view, comfort measures are, in most cases, linked to complexification, as complexity precisely solves the problems of a complex society.

With the development of digitization, the distance between an elite concentrating power and the rest of the population is amplified. This acceleration of inequalities is not a side effect but a presupposition on which the construction of the digital tool was based \cite{monnin_2022}. Illich addresses this phenomenon in the concept of polarization. This polarization refers to the separation of a knowledgeable elite that masters the tool and its users :"Under the thrust of the expanding mega-machine, the power to decide the fate of all becomes concentrated in the hands of a few. And, in this growth frenzy, innovations that improve the fortunes of the privileged minority grow even faster than the overall product" \cite{illich_1973}. This increase in the verticality of the system and the rise of an elite represent a complexification of society that becomes obstructive to political action. In such a highly hierarchical system, elites maintain the status quo, while the rest of the population has no influence on the political system what so ever. Complexification threatens individuals' right to voice, thereby obstructing any attempt of simplification of the society.

The digital de-escalation becomes impossible in a society where nobody has no control, where complexity maintains the inertia of development. The digitized complex system then becomes a driving force in its own right, with its will for growth.

This section has therefore identified three key points to analyze digital technologies development irreversibility. These technologies make people dependent both as individuals and as a society. Their network structure creates increasing returns that drive their progression, making the phenomenon of voluntary decline very unlikely. Finally, the monetary and political cost of dismantling appears excessively high compared to the short-term benefits it might bring. The following section therefore explores modes of action to allow individual to circumvent these systemic effects.

\section{Toward Simplification Perspectives: propositions}

The ratchet effects represent a real obstacle to a shift towards a paradigm of degrowth, simplification of society, and digital de-escalation. However, these effects are not insurmountable obstacles, and some societies in the past have already experienced a chosen reduction in their complexity\footnote{Yet, it is also important to mention that these ratchet effects are not the only obstacles to degrowth, and one should not believe that overcoming only these ratchet effects would necessarily lead to a new trajectory toward a decline of complexity.} \cite{tainter_2006}. 
Strategies can be implemented to initiate a shift in the digital usage, using trajectory of degrowth\footnote{One example is the concept of \textit{disintermediation} designs theorized by Raghavan, which aims to "remove intermediaries to simplify a system while simultaneously maintaining most or all of the system’s benefits" \cite{raghavan_2017}}.
 In this section, some perspectives are proposed to initiate a reflection on ways to transition towards a paradigm of degrowth, simplification, and digital de-escalation.

\subsection{Proposition to initiate Changes from the Bottom}

Having an impact on the ratchet effects that operate at the scale of the entire system appears highly challenging. A complex system such as the techno-industrial society in which we live seems to have its own expansionary will that exceeds the power of individuals, and having a hold on this system appears impossible \cite{illich_1973}. The most appropriate action seems therefore to be driven from the bottom-up: through communities, associations, networks of people who know each other and can carry out common projects, local and grassroots initiatives.

Radical transition thinkers usually emphasize the need for collective action on a small scale in various forms and for various reasons, but with the common point that this scale of action allows for leverage, a point of anchoring on which to act, and greater freedom to act by limiting the systemic barriers described above. At the end of the 20th century, Murray Bookchin proposed \textit{libertarian municipalism} \cite{bookchin_1990}, i.e., the end of the nation-state and the creation of a federation of self-governing municipalities, where people are connected by the daily life they share in a territory. This alternative, based on direct democracy and decentralization, aims to respond more pragmatically to the needs of individuals. 

However, as highlighted in an article by Kostakis et al. \cite{kostakis_2023}: "if degrowth is to contribute to global social/economic/political/ecological transformation, it must move past localism and dry critiques of socioeconomic metabolisms on more and wider scales, such as the regional and the global." To integrate both the systemic vision of a global degrowth transition while maintaining the transcendent and empowering nature of local action, the authors promote cosmolocalism: "the methods to bridge local communities in networks of shared resources and products." This societal model allows local communities to evolve as equals by sharing their knowledge and resources. This societal model promotes the \textit{economy of commons} and relies heavily on the notion of \textit{digital commons}. This cosmolocalist approach focuses on a vision of technical tools where the knowledge is shared, but the production is decentralized and adapted to local specifications. Through this production and societal model, local communities can overcome the obstacles for degrowth described earlier by reducing their dependence on global value chains, and by basing themselves on "the values of reciprocity and self-organisation that prioritise local autonomy and cultural diversity but also a sense of global common benefit." According to the article, the technical tools enabled and produced by this organizational model are compatible with degrowth: more durable tools that meet local needs, with more local materials, etc.

In the case of digitization, this change of production organization must correspond to technologies that are no longer imposed from the top but reappropriated and developed from the bottom. Numerous examples of digital projects at a local scale already exist. For example, the long-range Wi-Fi network named "StreetNet" developed in Cuba, "a grassroots information infrastructure that local residents developed in response to government internet access restrictions" \cite{sharma_2023}. Tomlinson discusses frugal innovation\footnote{"It consists in a form of improvisation, using the limited resources available at hand to provide low-cost and basic solutions"}, and emphasizes the localized nature of this practice: "usually done through homegrown solutions for local problems" \cite{tomlinson_2017}. De Valk emphasizes the benefits of this diversity in emerging small-scale digital forms, highlighting the importance of their networking : "This longitudinal view hopes to demonstrate the wealth this diversity of practices, and the thinking that informs them, brings. They can enrich each other and together provide a stronger counternarrative: alternatives are possible and already exist" \cite{devalk_2021}.

\subsection{Toward Renunciation: a Protocol}

The local and territorial scale in action, however, may still remain subject to an individual ratchet effect related to the attachments acquired to the direct \textit{subsistence network}. 
The shift from the bottom towards digital de-escalation therefore should necessarily take into account these attachments composed of both \textit{affective condensation} and \textit{dependence}. In addition, even in a perspective of shifting towards degrowth, individuals inherit a whole range of material and immaterial constraints left by the techno-industrial society, for instance pollutant infrastructures, disposed waste, addiction to smartphones, dependence to organizations, etc. This is what Landivar calls \textit{patrimonial continuity} : the heritage of a "capitalist world as patrimony, imposing itself on us" \cite{landivar_2021}. The author proposes to carry out protocols of \textit{ecological redirection} to concretely support the paradigm shift in ecology despite this heritage. In other worlds, these protocols aim to enable a radical ontological paradigm shift\footnote{which in the case of this article focused on digital complexity, can correspond to the transition from a growth paradigm to a degrowth paradigm} while taking into account our heritage.

These protocols consist first of mapping the attachments of individuals and understanding their \textit{subsistence networks}. Once this phase is established, the individuals concerned can collectively decide to renounce certain attachments \cite{monnin_2023} while reflecting on new means of subsistence that radically detach from the growth society. According to Tomlinson, support for renunciation is necessary, as he summarizes the dynamics of \textit{retreat of innovation} described in the literature: "while adoption of an innovation carries with it previously unknown benefits, the abandonment of an innovation is accompanied by known costs. Therefore, the process of abandonment is made under a condition of more consistent information than in the process of adoption, when the stakeholder first discovers the innovation" \cite{tomlinson_2017}. This highlights the importance of adapted support methods, that enable people to understand the benefits of renouncing to a digital technology for instance.

The \textit{ecological redirection} method allows for overcoming the barrier of attachments and proposing an alternative directed towards a horizon other than that of growth while redesigning the \textit{subsistence networks} that allow individuals to live decently. However, this perspective of \textit{ecological redirection} must be specified for the case of the digital realm through a set of methods adapted to its characteristics. This opens up a substantial field of research that can be explored in the digital case addressed in this paper.

\subsection{Shifting the Digital Realm}

Attachments operating in the digital realm are especially strong in terms of dependence, which challenges any renunciations processes, with the \textit{subsistence network} rupture risk. 
However, maintaining the digital status quo is undesirable in an objective focused on degrowth. Thus, \textit{ecological redirection} in the case of the digitization may need new forms of digital technologies. These new digital tools must help individuals and society to transition towards degrowth without completely breaking their \textit{subsistence networks}. They should facilitate a paradigm shift while ensuring the viability of the associated new way of life. Some examples of what could be considered such forms of digital technologies may already exist. The website https://solar.lowtechmagazine.com \cite{abbing_2021}, the MESH long-range Wi-Fi network in Detroit \cite{huguet_2022}, the "sneakernet" practices in India \cite{roussilhe_2020}, or the wireless networks enabling the exchange of information among nearby user terminals explored by Schmitt and Belding \cite{schmitt_2016} are relevant examples.

\section{Conclusion}
The contribution of knowledge on the social and bio-physical limits encountered by computing, as well as the proposal of new strategies to adapt the digital sector to these constraints, is an important and well-represented effort in the \textit{Computing Within Limits} community. For a post-growth digital future to emerge, this paper seek to understand the dynamics that accelerate its development and hinder its decline. The study of the link between computing, complexity and degrowth is valuable in this regard.

This article pointed out that the digital has a strong connection with complexity. Indeed, the development of infrastructural complexity linked to its form of \textit{Large Technological System} sustains the socio-political complexity of our society. Thanks to digitization, society grows, exchanges more information, diversifies in its social roles, etc. Conversely, this complexity supports the techno-industry that relentlessly develops the digital network. This strong intertwining between the digitization and complexity is geared towards growth.

On the other hand, some authors highlight an ontological complexity hindered by the grip of digital technologies in Western lives. These technologies deprive individuals of  relationships with nature and other humans, and confine society within a form of reductionist paradigm.

Since infrastructural and socio-political complexity is constantly increasing but the energy and material resources necessary for this growth are reaching their limits, current societies risk having to face collapse or controlled and democratically chosen simplification. The latter then becomes a goal for degrowth: the reduction of complexity to ensure resilience. However, this socio-political and infrastructural simplification inherent to degrowth cannot happen without an ontological complexification shift. 

As showed in this paper, the ratchet effects currently prevent this change of heading. These effects corresponding to irreversibility in the development of digital are behavioral (we are individually attached and dependent on digital), technological (the network expansion lock-in), and political (the need for legitimacy among elected officials prevents them from dismantling infrastructure, which would cost individuals without bringing short-term benefits).

Facing these last two systemic ratchet effects, it seems necessary to find a will for ecological redirection with a bottom-up approach. However, the behavioral ratchet effect remains an obstacle to overcome. Thus, this research invites to open to a new field of study: understanding individuals' attachments to the digital realm and finding means to renounce them (collectively in a democratic manner) while finding alternatives to ensure individuals subsistence. Methodologies and technical tools will need to be developed to support collectives of individuals in their transition towards digital de-escalation.

\bibliographystyle{ACM-Reference-Format}
\bibliography{sample-base}


\begin{thebibliography}{66}


\ifx \showCODEN    \undefined \def \showCODEN     #1{\unskip}     \fi
\ifx \showDOI      \undefined \def \showDOI       #1{#1}\fi
\ifx \showISBNx    \undefined \def \showISBNx     #1{\unskip}     \fi
\ifx \showISBNxiii \undefined \def \showISBNxiii  #1{\unskip}     \fi
\ifx \showISSN     \undefined \def \showISSN      #1{\unskip}     \fi
\ifx \showLCCN     \undefined \def \showLCCN      #1{\unskip}     \fi
\ifx \shownote     \undefined \def \shownote      #1{#1}          \fi
\ifx \showarticletitle \undefined \def \showarticletitle #1{#1}   \fi
\ifx \showURL      \undefined \def \showURL       {\relax}        \fi
\providecommand\bibfield[2]{#2}
\providecommand\bibinfo[2]{#2}
\providecommand\natexlab[1]{#1}
\providecommand\showeprint[2][]{arXiv:#2}

\bibitem[Arthur(1994)]%
        {arthur_1994}
\bibfield{author}{\bibinfo{person}{W.~Brian Arthur}.} \bibinfo{year}{1994}\natexlab{}.
\newblock \bibinfo{booktitle}{\emph{Increasing returns and path dependence in the economy}}.
\newblock \bibinfo{publisher}{University of Michigan Press}, \bibinfo{address}{Ann Arbor}.
\newblock
\showISBNx{978-0-472-09496-7 978-0-472-06496-0}


\bibitem[Barrau and Guilbaud(2022)]%
        {barrau_2022}
\bibfield{author}{\bibinfo{person}{Aurélien Barrau} {and} \bibinfo{person}{Carole Guilbaud}.} \bibinfo{year}{2022}\natexlab{}.
\newblock \bibinfo{booktitle}{\emph{Il faut une révolution politique, poétique et philosophique}}.
\newblock \bibinfo{publisher}{Éditions Zulma}, \bibinfo{address}{Paris}.
\newblock
\showISBNx{979-10-387-0129-8}


\bibitem[Bihouix(2021)]%
        {bihouix_2021}
\bibfield{author}{\bibinfo{person}{Philippe Bihouix}.} \bibinfo{year}{2021}\natexlab{}.
\newblock \bibinfo{booktitle}{\emph{L'âge des low tech: vers une civilisation techniquement soutenable}}.
\newblock \bibinfo{publisher}{Éditions du Seuil}, \bibinfo{address}{Paris}.
\newblock
\showISBNx{978-2-7578-8951-0}


\bibitem[Bihouix and de~Guillebon(2010)]%
        {bihouix_2010}
\bibfield{author}{\bibinfo{person}{Philippe Bihouix} {and} \bibinfo{person}{Benoît de Guillebon}.} \bibinfo{year}{2010}\natexlab{}.
\newblock \bibinfo{booktitle}{\emph{Quel futur pour les métaux ? {Raréfaction} des métaux, un nouveau défi pour la société}}.
\newblock \bibinfo{publisher}{EDP Sciences}, \bibinfo{address}{Les Ulis, france}.
\newblock
\showISBNx{978-2-7598-0549-5}


\bibitem[Blessing(2009)]%
        {DRM}
\bibfield{author}{\bibinfo{person}{Lucienne~TM Blessing}.} \bibinfo{year}{2009}\natexlab{}.
\newblock \bibinfo{booktitle}{\emph{{DRM}, a {Design} {Research} {Methodology}} (\bibinfo{edition}{1st ed. 2009.} ed.)}.
\newblock \bibinfo{publisher}{Springer London}, \bibinfo{address}{London}.
\newblock
\showISBNx{978-1-282-28853-9}
\urldef\tempurl%
\url{https://doi.org/10.1007/978-1-84882-587-1}
\showDOI{\tempurl}


\bibitem[Boluda et~al\mbox{.}(2021)]%
        {boluda_2021}
\bibfield{author}{\bibinfo{person}{In{\` e}s~Moreno Boluda}, \bibinfo{person}{Elizabeth Patitsas}, {and} \bibinfo{person}{Peter McMahan}.} \bibinfo{year}{2021}\natexlab{}.
\newblock \showarticletitle{What do {Computer} {Scientists} {Know} {About} {Conflict} {Minerals}?}. In \bibinfo{booktitle}{\emph{Seventh {Workshop} on {Computing} within {Limits} 2021}}. LIMITS.
\newblock
\newblock
\shownote{https://limits.pubpub.org/pub/1u45epgg}.


\bibitem[{Bonnet Emmanuel} et~al\mbox{.}(2021)]%
        {landivar_2021}
\bibfield{author}{\bibinfo{person}{{Bonnet Emmanuel}}, \bibinfo{person}{{Landivar Diego}}, {and} \bibinfo{person}{{Monnin Alexandre}}.} \bibinfo{year}{2021}\natexlab{}.
\newblock \bibinfo{booktitle}{\emph{Héritage et fermeture: une écologie du démantèlement}}.
\newblock \bibinfo{publisher}{éditions divergences}, \bibinfo{address}{Paris}.
\newblock
\showISBNx{979-10-97088-37-8}


\bibitem[Bookchin(1990)]%
        {bookchin_1990}
\bibfield{author}{\bibinfo{person}{Murray Bookchin}.} \bibinfo{year}{1990}\natexlab{}.
\newblock \bibinfo{booktitle}{\emph{Remaking society} (\bibinfo{edition}{2. print} ed.)}.
\newblock Number 121 in \bibinfo{series}{Black {Rose} books}. \bibinfo{publisher}{Black Rose Books}, \bibinfo{address}{Montréal}.
\newblock
\showISBNx{978-0-921689-02-7 978-0-921689-03-4}


\bibitem[{Bordage Frédéric}(2019)]%
        {bordage_2019}
\bibfield{author}{\bibinfo{person}{{Bordage Frédéric}}.} \bibinfo{year}{2019}\natexlab{}.
\newblock \bibinfo{booktitle}{\emph{Sobriété numérique: les clés pour agir}}.
\newblock \bibinfo{publisher}{Buchet-Chastel}, \bibinfo{address}{Paris}.
\newblock
\showISBNx{978-2-283-03215-2}


\bibitem[Bremer et~al\mbox{.}(2022)]%
        {bremer_2022}
\bibfield{author}{\bibinfo{person}{Christina Bremer}, \bibinfo{person}{Bran Knowles}, {and} \bibinfo{person}{Adrian Friday}.} \bibinfo{year}{2022}\natexlab{}.
\newblock \showarticletitle{Have {We} {Taken} {On} {Too} {Much}?: {A} {Critical} {Review} of the {Sustainable} {HCI} {Landscape}}. In \bibinfo{booktitle}{\emph{Proceedings of the 2022 {CHI} {Conference} on {Human} {Factors} in {Computing} {Systems}}} \emph{(\bibinfo{series}{{CHI} '22})}. \bibinfo{publisher}{Association for Computing Machinery}, \bibinfo{address}{New York, NY, USA}, \bibinfo{pages}{1--11}.
\newblock
\showISBNx{978-1-4503-9157-3}
\urldef\tempurl%
\url{https://doi.org/10.1145/3491102.3517609}
\showDOI{\tempurl}


\bibitem[Chabanne et~al\mbox{.}(2023)]%
        {chabanne_2023}
\bibfield{author}{\bibinfo{person}{Simon Chabanne}, \bibinfo{person}{Romain Couillet}, \bibinfo{person}{Céleste De~Bourmont}, \bibinfo{person}{Pierre-Thomas Demars}, \bibinfo{person}{Valentin Girard}, \bibinfo{person}{Sacha Hodencq}, \bibinfo{person}{Julie Mignerey-Koelsch}, {and} \bibinfo{person}{Grégoire Poissonnier}.} \bibinfo{year}{2023}\natexlab{}.
\newblock \showarticletitle{Les impacts sociaux du numérique, grands oubliés de la transition écologique?} \bibinfo{address}{Grenoble}.
\newblock
\urldef\tempurl%
\url{https://polaris.imag.fr/romain.couillet/docs/articles/gretsi_social.pdf}
\showURL{%
\tempurl}


\bibitem[Citton(2022)]%
        {citton_2022}
\bibfield{author}{\bibinfo{person}{Yves Citton}.} \bibinfo{year}{2022}\natexlab{}.
\newblock \showarticletitle{Smartness de surveillance et intelligences d’improvisation: deux écologies du smartphone}.
\newblock In \bibinfo{booktitle}{\emph{Écologies du smartphone}}. \bibinfo{publisher}{Le bord de l'eau}, \bibinfo{address}{Lormond}.
\newblock
\showISBNx{978-2-35687-791-8}


\bibitem[Comber and Eriksson(2023)]%
        {comber_2023}
\bibfield{author}{\bibinfo{person}{Rob Comber} {and} \bibinfo{person}{Elina Eriksson}.} \bibinfo{year}{2023}\natexlab{}.
\newblock \showarticletitle{Computing as {Ecocide}}. In \bibinfo{booktitle}{\emph{Ninth {Computing} within {Limits} 2023}}. \bibinfo{publisher}{LIMITS}, \bibinfo{address}{Virtual}.
\newblock
\urldef\tempurl%
\url{https://doi.org/10.21428/bf6fb269.9fcdd0c0}
\showDOI{\tempurl}


\bibitem[Couillet and Poissonnier(2023)]%
        {couillet_2023}
\bibfield{author}{\bibinfo{person}{Romain Couillet} {and} \bibinfo{person}{Grégoire Poissonnier}.} \bibinfo{year}{2023}\natexlab{}.
\newblock \showarticletitle{Pourquoi et comment démanteler le numérique?}
\newblock
\urldef\tempurl%
\url{https://polaris.imag.fr/romain.couillet/docs/articles/gretsi_demantelement.pdf}
\showURL{%
\tempurl}


\bibitem[de~la Porte and Laigle(2024)]%
        {france_inter}
\bibfield{author}{\bibinfo{person}{Xavier de~la Porte} {and} \bibinfo{person}{Fabrice Laigle}.} \bibinfo{year}{2024}\natexlab{}.
\newblock \bibinfo{title}{Une vie sans {Internet}}.
\newblock
\newblock
\urldef\tempurl%
\url{https://www.radiofrance.fr/franceinter/podcasts/le-code-a-change/le-code-a-change-5-5175902}
\showURL{%
\tempurl}
\newblock
\shownote{Section: Société}.


\bibitem[De~Valk(2021)]%
        {devalk_2021}
\bibfield{author}{\bibinfo{person}{Marloes De~Valk}.} \bibinfo{year}{2021}\natexlab{}.
\newblock \showarticletitle{A pluriverse of local worlds: {A} review of {Computing} within {Limits} related terminology and practices}. In \bibinfo{booktitle}{\emph{{LIMITS} {Workshop} on {Computing} within {Limits}}}. \bibinfo{publisher}{PubPub}.
\newblock
\urldef\tempurl%
\url{https://assets.pubpub.org/eg52sqwg/51623536420052.pdf}
\showURL{%
\tempurl}


\bibitem[Descola(2005)]%
        {descola_2005}
\bibfield{author}{\bibinfo{person}{Philippe Descola}.} \bibinfo{year}{2005}\natexlab{}.
\newblock \bibinfo{booktitle}{\emph{Par-delà nature et culture}}.
\newblock \bibinfo{publisher}{NRF : Gallimard}, \bibinfo{address}{Paris?}
\newblock
\showISBNx{978-2-07-077263-6}


\bibitem[{Duesenberry James Stemble}(1962)]%
        {duesenberry_1962}
\bibfield{author}{\bibinfo{person}{{Duesenberry James Stemble}}.} \bibinfo{year}{1962}\natexlab{}.
\newblock \bibinfo{booktitle}{\emph{Income saving and the theory of consumer behavior}}.
\newblock \bibinfo{publisher}{Harvard University Press}, \bibinfo{address}{Cambridge}.
\newblock


\bibitem[España et~al\mbox{.}(2023)]%
        {espana_2023}
\bibfield{author}{\bibinfo{person}{Sergio España}, \bibinfo{person}{Willem Hulst}, \bibinfo{person}{Nivard Jansen}, {and} \bibinfo{person}{Daniel Pargmam}.} \bibinfo{year}{2023}\natexlab{}.
\newblock \showarticletitle{Untangling the {Relationship} {Between} {Degrowth} and {ICT}}. In \bibinfo{booktitle}{\emph{2023 {International} {Conference} on {ICT} for {Sustainability} ({ICT4S})}}. \bibinfo{pages}{1--12}.
\newblock
\urldef\tempurl%
\url{https://doi.org/10.1109/ICT4S58814.2023.00010}
\showDOI{\tempurl}


\bibitem[Ferreboeuf(2018)]%
        {lean_2018}
\bibfield{author}{\bibinfo{person}{Hugues Ferreboeuf}.} \bibinfo{year}{2018}\natexlab{}.
\newblock \bibinfo{booktitle}{\emph{Lean {ICT} : {Pour} une sobriété numérique}}.
\newblock \bibinfo{type}{{T}echnical {R}eport}. \bibinfo{institution}{The Shift Project}.
\newblock


\bibitem[Forti(2020)]%
        {forti_2020}
\bibfield{author}{\bibinfo{person}{Vanessa Forti}.} \bibinfo{year}{2020}\natexlab{}.
\newblock \bibinfo{booktitle}{\emph{The {Global} {E}-waste {Monitor} 2020}}.
\newblock \bibinfo{type}{{T}echnical {R}eport}. \bibinfo{institution}{UNITAR}.
\newblock


\bibitem[Geldron(2017)]%
        {geldron_2017}
\bibfield{author}{\bibinfo{person}{Alain Geldron}.} \bibinfo{year}{2017}\natexlab{}.
\newblock \bibinfo{booktitle}{\emph{L'épuisement des métaux et minéraux : faut-il s'inquiéter ?}}
\newblock


\bibitem[Georgescu-Roegen(1971)]%
        {georgescu_1971}
\bibfield{author}{\bibinfo{person}{Nicholas Georgescu-Roegen}.} \bibinfo{year}{1971}\natexlab{}.
\newblock \bibinfo{booktitle}{\emph{The entropy law and the economic process}}.
\newblock \bibinfo{publisher}{Harvard University Press}, \bibinfo{address}{Cambridge, Mass}.
\newblock
\showISBNx{978-0-674-25780-1}


\bibitem[Gisondon et~al\mbox{.}(2020)]%
        {gisondon_2020}
\bibfield{author}{\bibinfo{person}{Skyler Gisondon}, \bibinfo{person}{Kara Hayward}, {and} \bibinfo{person}{Vincent Kartheiser}.} \bibinfo{year}{2020}\natexlab{}.
\newblock \bibinfo{title}{The Social Dilemma}.
\newblock
\newblock


\bibitem[Hickel(2020)]%
        {hickel_2020}
\bibfield{author}{\bibinfo{person}{Jason Hickel}.} \bibinfo{year}{2020}\natexlab{}.
\newblock \bibinfo{booktitle}{\emph{Less is more: how degrowth will save the world}}.
\newblock \bibinfo{publisher}{William Heinemann}, \bibinfo{address}{London}.
\newblock
\showISBNx{978-1-78515-249-8 978-1-78515-250-4}


\bibitem[Huguet(2022)]%
        {huguet_2022}
\bibfield{author}{\bibinfo{person}{François Huguet}.} \bibinfo{year}{2022}\natexlab{}.
\newblock \showarticletitle{Jeux de ficelles sans-fil à {Motown}. {Les} réseaux {MESH} de {Détroit} comme formes de lyannajismes numériques.}
\newblock In \bibinfo{booktitle}{\emph{Écologies du smartphone}}. \bibinfo{publisher}{Le bord de l'eau}, \bibinfo{address}{Lormond}.
\newblock
\showISBNx{978-2-35687-791-8}


\bibitem[Illich(1973)]%
        {illich_1973}
\bibfield{author}{\bibinfo{person}{Ivan Illich}.} \bibinfo{year}{1973}\natexlab{}.
\newblock \bibinfo{booktitle}{\emph{Tools for conviviality} (\bibinfo{edition}{american first edition} ed.)}.
\newblock \bibinfo{publisher}{Calder and Boyars}.
\newblock
\showISBNx{978-0-7145-0973-0}


\bibitem[Izoard(2020)]%
        {izoard_2020}
\bibfield{author}{\bibinfo{person}{Celia Izoard}.} \bibinfo{year}{2020}\natexlab{}.
\newblock \showarticletitle{Les réalités occultées du « progrès » technique : inégalités et désastres socio-écologiques}.
\newblock In \bibinfo{booktitle}{\emph{Low-tech : face au tout numérique, se réapproprier les technologies} (\bibinfo{edition}{ritimo} ed.)}. \bibinfo{pages}{27--33}.
\newblock
\showISBNx{978-2-914180-87-0}


\bibitem[Izoard(2024)]%
        {izoard_2024}
\bibfield{author}{\bibinfo{person}{Celia Izoard}.} \bibinfo{year}{2024}\natexlab{}.
\newblock \bibinfo{booktitle}{\emph{La ruée minière au {XXIe} siècle: enquête sur les métaux à l'ère de la transition}}.
\newblock \bibinfo{publisher}{Points}, \bibinfo{address}{Paris}.
\newblock
\showISBNx{978-2-02-151528-2}


\bibitem[Jackson(2011)]%
        {jackson_2011}
\bibfield{author}{\bibinfo{person}{Tim Jackson}.} \bibinfo{year}{2011}\natexlab{}.
\newblock \bibinfo{booktitle}{\emph{Prosperity without growth: economics for a finite planet} (\bibinfo{edition}{pbk. ed} ed.)}.
\newblock \bibinfo{publisher}{Earthscan}, \bibinfo{address}{London ; Washington, DC}.
\newblock
\showISBNx{978-1-84971-323-8}


\bibitem[Jang et~al\mbox{.}(2017)]%
        {jang_2017}
\bibfield{author}{\bibinfo{person}{Esther Jang}, \bibinfo{person}{Matthew Johnson}, \bibinfo{person}{Edward Burnell}, {and} \bibinfo{person}{Kurtis Heimerl}.} \bibinfo{year}{2017}\natexlab{}.
\newblock \showarticletitle{Unplanned {Obsolescence}: {Hardware} and {Software} {After} {Collapse}}. In \bibinfo{booktitle}{\emph{Proceedings of the 2017 {Workshop} on {Computing} {Within} {Limits}}} \emph{(\bibinfo{series}{{LIMITS} '17})}. \bibinfo{publisher}{Association for Computing Machinery}, \bibinfo{address}{New York, NY, USA}, \bibinfo{pages}{93--101}.
\newblock
\showISBNx{978-1-4503-4950-5}
\urldef\tempurl%
\url{https://doi.org/10.1145/3080556.3080566}
\showDOI{\tempurl}


\bibitem[Kallis et~al\mbox{.}(2020)]%
        {kallis_2020}
\bibfield{author}{\bibinfo{person}{Giorgos Kallis}, \bibinfo{person}{Susan Paulson}, \bibinfo{person}{Giacomo D'Alisa}, {and} \bibinfo{person}{Federico Demaria}.} \bibinfo{year}{2020}\natexlab{}.
\newblock \bibinfo{booktitle}{\emph{The case for degrowth}}.
\newblock \bibinfo{publisher}{Polity}, \bibinfo{address}{Cambridge, UK Medford, MA}.
\newblock
\showISBNx{978-1-5095-3562-0 978-1-5095-3563-7}


\bibitem[Kerschner et~al\mbox{.}(2018)]%
        {kerschner_2018}
\bibfield{author}{\bibinfo{person}{Christian Kerschner}, \bibinfo{person}{Petra Wächter}, \bibinfo{person}{Linda Nierling}, {and} \bibinfo{person}{Melf-Hinrich Ehlers}.} \bibinfo{year}{2018}\natexlab{}.
\newblock \showarticletitle{Degrowth and {Technology}: {Towards} feasible, viable, appropriate and convivial imaginaries}.
\newblock \bibinfo{journal}{\emph{Journal of Cleaner Production}}  \bibinfo{volume}{197} (\bibinfo{date}{Oct.} \bibinfo{year}{2018}), \bibinfo{pages}{1619--1636}.
\newblock
\showISSN{0959-6526}
\urldef\tempurl%
\url{https://doi.org/10.1016/j.jclepro.2018.07.147}
\showDOI{\tempurl}


\bibitem[Kostakis et~al\mbox{.}(2023)]%
        {kostakis_2023}
\bibfield{author}{\bibinfo{person}{Vasilis Kostakis}, \bibinfo{person}{Vasilis Niaros}, {and} \bibinfo{person}{Chris Giotitsas}.} \bibinfo{year}{2023}\natexlab{}.
\newblock \showarticletitle{Beyond global versus local: illuminating a cosmolocal framework for convivial technology development}.
\newblock \bibinfo{journal}{\emph{Sustainability Science}} \bibinfo{volume}{18}, \bibinfo{number}{5} (\bibinfo{date}{Sept.} \bibinfo{year}{2023}), \bibinfo{pages}{2309--2322}.
\newblock
\showISSN{1862-4057}
\urldef\tempurl%
\url{https://doi.org/10.1007/s11625-023-01378-1}
\showDOI{\tempurl}


\bibitem[Latouche(2006)]%
        {latouche_2006}
\bibfield{author}{\bibinfo{person}{Serge Latouche}.} \bibinfo{year}{2006}\natexlab{}.
\newblock \bibinfo{booktitle}{\emph{Le pari de la décroissance}}.
\newblock \bibinfo{publisher}{Fayard}, \bibinfo{address}{Paris}.
\newblock
\showISBNx{978-2-213-62914-8}


\bibitem[{Leonarduzzi Inès}(2021)]%
        {leonarduzzi_2021}
\bibfield{author}{\bibinfo{person}{{Leonarduzzi Inès}}.} \bibinfo{year}{2021}\natexlab{}.
\newblock \bibinfo{booktitle}{\emph{Réparer le futur: du numérique à l'écologie}}.
\newblock \bibinfo{publisher}{Éditions de l'Observatoire}, \bibinfo{address}{Paris}.
\newblock
\showISBNx{979-10-329-1616-2}


\bibitem[Liebowitz and Margolis(1995)]%
        {liebowitz_1995}
\bibfield{author}{\bibinfo{person}{S.~J. Liebowitz} {and} \bibinfo{person}{Stephen~E. Margolis}.} \bibinfo{year}{1995}\natexlab{}.
\newblock \showarticletitle{Path {Dependence}, {Lock}-in, and {History}}.
\newblock \bibinfo{journal}{\emph{Journal of Law, Economics, \& Organization}} \bibinfo{volume}{11}, \bibinfo{number}{1} (\bibinfo{year}{1995}), \bibinfo{pages}{205--226}.
\newblock
\showISSN{8756-6222}
\urldef\tempurl%
\url{https://www.jstor.org/stable/765077}
\showURL{%
\tempurl}
\newblock
\shownote{Publisher: Oxford University Press}.


\bibitem[{Lievens Laurent}(2022)]%
        {lievens_2022}
\bibfield{author}{\bibinfo{person}{{Lievens Laurent}}.} \bibinfo{year}{2022}\natexlab{}.
\newblock \bibinfo{booktitle}{\emph{Décroissance et néodécroissance. {L}'engagement militant pour sortir de l'économisme écocidaire}}.
\newblock \bibinfo{publisher}{Presses Universitaires}, \bibinfo{address}{Louvain-La-Neuve}.
\newblock
\showISBNx{978-2-39061-244-5}


\bibitem[Lopez(2022)]%
        {lopez_2022}
\bibfield{author}{\bibinfo{person}{Fanny Lopez}.} \bibinfo{year}{2022}\natexlab{}.
\newblock \bibinfo{booktitle}{\emph{À bout de flux}}.
\newblock \bibinfo{publisher}{Editions Divergences}.
\newblock
\showISBNx{979-10-97088-50-7}
\newblock
\shownote{Country: FR 20 cm.}.


\bibitem[Louv(2008)]%
        {louv_2008}
\bibfield{author}{\bibinfo{person}{Richard Louv}.} \bibinfo{year}{2008}\natexlab{}.
\newblock \bibinfo{booktitle}{\emph{Last child in the woods: saving our children from nature-deficit disorder} (\bibinfo{edition}{updated and expanded} ed.)}.
\newblock \bibinfo{publisher}{Algonquin Books of Chapel Hill}, \bibinfo{address}{Chapel Hill, N.C}.
\newblock
\showISBNx{978-1-56512-605-3}
\newblock
\shownote{OCLC: ocn183879632}.


\bibitem[Meadows(1999)]%
        {meadows_1999}
\bibfield{author}{\bibinfo{person}{Donella~H. Meadows}.} \bibinfo{year}{1999}\natexlab{}.
\newblock \bibinfo{booktitle}{\emph{Leverage {Points}: {Places} to {Intervene} in a {System}}}.
\newblock \bibinfo{type}{{T}echnical {R}eport}. \bibinfo{institution}{The Sustainability Institute}.
\newblock
\urldef\tempurl%
\url{https://1a0c26.p3cdn2.secureserver.net/wp-content/userfiles/Leverage_Points.pdf}
\showURL{%
\tempurl}


\bibitem[Meadows and {Club of Rome}(1972)]%
        {meadows_1972}
\bibfield{editor}{\bibinfo{person}{Donella~H. Meadows} {and} \bibinfo{person}{{Club of Rome}}} (Eds.). \bibinfo{year}{1972}\natexlab{}.
\newblock \bibinfo{booktitle}{\emph{The {Limits} to growth: a report for the {Club} of {Rome}'s project on the predicament of mankind}}.
\newblock \bibinfo{publisher}{Universe Books}, \bibinfo{address}{New York}.
\newblock
\showISBNx{978-0-87663-165-2}


\bibitem[Mies(1986)]%
        {mies_1986}
\bibfield{author}{\bibinfo{person}{Maria Mies}.} \bibinfo{year}{1986}\natexlab{}.
\newblock \bibinfo{booktitle}{\emph{Patriarchy and accumulation on a world scale: women in the international division of labour}}.
\newblock \bibinfo{publisher}{Zed Books Distributed in the U.S.A. and Canada by Humanities Press}, \bibinfo{address}{London Atlantic Highlands, N.J., USA Atlantic Highlands, N.J}.
\newblock
\showISBNx{978-0-86232-341-7}


\bibitem[Miller(2022)]%
        {miller_2022}
\bibfield{author}{\bibinfo{person}{Chris Miller}.} \bibinfo{year}{2022}\natexlab{}.
\newblock \bibinfo{booktitle}{\emph{Chip war: the fight for the world's most critical technology} (\bibinfo{edition}{first scribner hardcover edition} ed.)}.
\newblock \bibinfo{publisher}{Scribner, an imprint of Simon \& Schuster}, \bibinfo{address}{New York}.
\newblock
\showISBNx{978-1-982172-00-8}
\newblock
\shownote{OCLC: on1296942188}.


\bibitem[Monnin(2022)]%
        {monnin_2022}
\bibfield{author}{\bibinfo{person}{Alexandre Monnin}.} \bibinfo{year}{2022}\natexlab{}.
\newblock \showarticletitle{Le numérique comme nouveau processus de biosphérisation}.
\newblock In \bibinfo{booktitle}{\emph{Écologies du smartphone}}. \bibinfo{publisher}{Le bord de l'eau}, \bibinfo{address}{Lormond}.
\newblock
\showISBNx{978-2-35687-791-8}


\bibitem[Monnin et~al\mbox{.}(2020)]%
        {monnin_2020}
\bibfield{author}{\bibinfo{person}{Alexandre Monnin}, \bibinfo{person}{José Halloy}, {and} \bibinfo{person}{Nicolas Nova}.} \bibinfo{year}{2020}\natexlab{}.
\newblock \showarticletitle{Au-delà du low tech: technologies zombies, soutenabilité et inventions}.
\newblock In \bibinfo{booktitle}{\emph{Low-tech : face au tout numérique, se réapproprier les technologies} (\bibinfo{edition}{ritimo} ed.)}. \bibinfo{pages}{120--128}.
\newblock
\showISBNx{978-2-914180-87-0}


\bibitem[{Monnin Alexandre}(2023)]%
        {monnin_2023}
\bibfield{author}{\bibinfo{person}{{Monnin Alexandre}}.} \bibinfo{year}{2023}\natexlab{}.
\newblock \bibinfo{booktitle}{\emph{Politiser le renoncement}}.
\newblock \bibinfo{publisher}{Éditions divergences}, \bibinfo{address}{Paris}.
\newblock
\showISBNx{979-10-97088-53-8}


\bibitem[Morin(1991)]%
        {morin_1991}
\bibfield{author}{\bibinfo{person}{Edgar Morin}.} \bibinfo{year}{1991}\natexlab{}.
\newblock \bibinfo{booktitle}{\emph{Les idées: leur habitat, leur vie, leurs mœurs, leur organisation}}.
\newblock Number~4 in \bibinfo{series}{La {Méthode}}. \bibinfo{publisher}{Editions du Seuil}, \bibinfo{address}{Paris}.
\newblock
\showISBNx{978-2-02-013669-3 978-2-02-005638-0}


\bibitem[Nardi et~al\mbox{.}(2018)]%
        {nardi_2018}
\bibfield{author}{\bibinfo{person}{Bonnie Nardi}, \bibinfo{person}{Bill Tomlinson}, \bibinfo{person}{Donald~J. Patterson}, \bibinfo{person}{Jay Chen}, \bibinfo{person}{Daniel Pargman}, \bibinfo{person}{Barath Raghavan}, {and} \bibinfo{person}{Birgit Penzenstadler}.} \bibinfo{year}{2018}\natexlab{}.
\newblock \showarticletitle{Computing within limits}.
\newblock \bibinfo{journal}{\emph{Commun. ACM}} \bibinfo{volume}{61}, \bibinfo{number}{10} (\bibinfo{date}{Sept.} \bibinfo{year}{2018}), \bibinfo{pages}{86--93}.
\newblock
\showISSN{0001-0782, 1557-7317}
\urldef\tempurl%
\url{https://doi.org/10.1145/3183582}
\showDOI{\tempurl}


\bibitem[{Parrique Timothée}(2022)]%
        {parrique_2022}
\bibfield{author}{\bibinfo{person}{{Parrique Timothée}}.} \bibinfo{year}{2022}\natexlab{}.
\newblock \bibinfo{booktitle}{\emph{Ralentir ou périr: l'économie de la décroissance}}.
\newblock \bibinfo{publisher}{Éditions du Seuil}, \bibinfo{address}{Paris}.
\newblock
\showISBNx{978-2-02-150809-3}


\bibitem[{Pitron Guillaume}(2021)]%
        {pitron_2021}
\bibfield{author}{\bibinfo{person}{{Pitron Guillaume}}.} \bibinfo{year}{2021}\natexlab{}.
\newblock \bibinfo{booktitle}{\emph{L'enfer numérique: voyage au bout d'un like}}.
\newblock \bibinfo{publisher}{Éditions Les liens qui Libèrent}, \bibinfo{address}{Paris}.
\newblock
\showISBNx{979-10-209-0996-1}


\bibitem[Rabhi(2010)]%
        {rabhi_2010}
\bibfield{author}{\bibinfo{person}{Pierre Rabhi}.} \bibinfo{year}{2010}\natexlab{}.
\newblock \bibinfo{booktitle}{\emph{Vers la sobriété heureuse} (\bibinfo{edition}{1re éd} ed.)}.
\newblock \bibinfo{publisher}{Actes sud}, \bibinfo{address}{Arles}.
\newblock
\showISBNx{978-2-7427-8967-2}


\bibitem[Raghavan and Pargman(2017)]%
        {raghavan_2017}
\bibfield{author}{\bibinfo{person}{Barath Raghavan} {and} \bibinfo{person}{Daniel Pargman}.} \bibinfo{year}{2017}\natexlab{}.
\newblock \showarticletitle{Means and {Ends} in {Human}-{Computer} {Interaction}: {Sustainability} through {Disintermediation}}. In \bibinfo{booktitle}{\emph{Proceedings of the 2017 {CHI} {Conference} on {Human} {Factors} in {Computing} {Systems}}} \emph{(\bibinfo{series}{{CHI} '17})}. \bibinfo{publisher}{Association for Computing Machinery}, \bibinfo{address}{New York, NY, USA}, \bibinfo{pages}{786--796}.
\newblock
\showISBNx{978-1-4503-4655-9}
\urldef\tempurl%
\url{https://doi.org/10.1145/3025453.3025542}
\showDOI{\tempurl}


\bibitem[Reporterre(2021)]%
        {reporterre}
\bibfield{author}{\bibinfo{person}{Reporterre}.} \bibinfo{year}{2021}\natexlab{}.
\newblock \bibinfo{title}{À {Taïwan}, la sécheresse menace la production de puces électroniques}.
\newblock
\newblock
\urldef\tempurl%
\url{https://reporterre.net/A-Taiwan-la-secheresse-menace-la-production-de-puces-electroniques}
\showURL{%
\tempurl}


\bibitem[Richardson et~al\mbox{.}(2023)]%
        {richardson_2023}
\bibfield{author}{\bibinfo{person}{Katherine Richardson}, \bibinfo{person}{Will Steffen}, \bibinfo{person}{Wolfgang Lucht}, \bibinfo{person}{Jørgen Bendtsen}, \bibinfo{person}{Sarah~E. Cornell}, \bibinfo{person}{Jonathan~F. Donges}, \bibinfo{person}{Markus Drüke}, \bibinfo{person}{Ingo Fetzer}, \bibinfo{person}{Govindasamy Bala}, \bibinfo{person}{Werner Von~Bloh}, \bibinfo{person}{Georg Feulner}, \bibinfo{person}{Stephanie Fiedler}, \bibinfo{person}{Dieter Gerten}, \bibinfo{person}{Tom Gleeson}, \bibinfo{person}{Matthias Hofmann}, \bibinfo{person}{Willem Huiskamp}, \bibinfo{person}{Matti Kummu}, \bibinfo{person}{Chinchu Mohan}, \bibinfo{person}{David Nogués-Bravo}, \bibinfo{person}{Stefan Petri}, \bibinfo{person}{Miina Porkka}, \bibinfo{person}{Stefan Rahmstorf}, \bibinfo{person}{Sibyll Schaphoff}, \bibinfo{person}{Kirsten Thonicke}, \bibinfo{person}{Arne Tobian}, \bibinfo{person}{Vili Virkki}, \bibinfo{person}{Lan Wang-Erlandsson}, \bibinfo{person}{Lisa Weber}, {and} \bibinfo{person}{Johan Rockström}.}
  \bibinfo{year}{2023}\natexlab{}.
\newblock \showarticletitle{Earth beyond six of nine planetary boundaries}.
\newblock \bibinfo{journal}{\emph{Science Advances}} \bibinfo{volume}{9}, \bibinfo{number}{37} (\bibinfo{date}{Sept.} \bibinfo{year}{2023}), \bibinfo{pages}{eadh2458}.
\newblock
\showISSN{2375-2548}
\urldef\tempurl%
\url{https://doi.org/10.1126/sciadv.adh2458}
\showDOI{\tempurl}


\bibitem[Rockström(2009)]%
        {rockstrom_2009}
\bibfield{author}{\bibinfo{person}{Johan Rockström}.} \bibinfo{year}{2009}\natexlab{}.
\newblock \showarticletitle{A safe operating space for humanity}.
\newblock \bibinfo{journal}{\emph{Nature}}  \bibinfo{volume}{461} (\bibinfo{date}{Sept.} \bibinfo{year}{2009}), \bibinfo{pages}{472--475}.
\newblock
\showISSN{00280836}
\urldef\tempurl%
\url{https://doi.org/10.1038/461472a}
\showDOI{\tempurl}
\newblock
\shownote{Publisher: Springer Nature}.


\bibitem[Roscam~Abbing(2021)]%
        {abbing_2021}
\bibfield{author}{\bibinfo{person}{Roel Roscam~Abbing}.} \bibinfo{year}{2021}\natexlab{}.
\newblock \showarticletitle{‘{This} is a solar-powered website, which means it sometimes goes offline’ : a design inquiry into degrowth and {ICT}}. \bibinfo{publisher}{PubPub}.
\newblock
\urldef\tempurl%
\url{https://urn.kb.se/resolve?urn=urn:nbn:se:mau:diva-55224}
\showURL{%
\tempurl}


\bibitem[Roussilhe(2020)]%
        {roussilhe_2020}
\bibfield{author}{\bibinfo{person}{Gauthier Roussilhe}.} \bibinfo{year}{2020}\natexlab{}.
\newblock \bibinfo{title}{Situer le numérique}.
\newblock
\newblock
\urldef\tempurl%
\url{https://gauthierroussilhe.com/ressources/situer-le-numerique}
\showURL{%
\tempurl}


\bibitem[Roussilhe and Monnin(2023)]%
        {podcast}
\bibfield{author}{\bibinfo{person}{Gauthier Roussilhe} {and} \bibinfo{person}{Alexandre Monnin}.} \bibinfo{year}{2023}\natexlab{}.
\newblock \bibinfo{title}{Défaire nos attachements à la numérisation ?}
\newblock
\newblock
\urldef\tempurl%
\url{https://open.spotify.com/show/7t9MOjGJg6sqY3yxzk4tYs}
\showURL{%
\tempurl}


\bibitem[Schmitt and Belding(2016)]%
        {schmitt_2016}
\bibfield{author}{\bibinfo{person}{Paul Schmitt} {and} \bibinfo{person}{Elizabeth Belding}.} \bibinfo{year}{2016}\natexlab{}.
\newblock \showarticletitle{Navigating connectivity in reduced infrastructure environments}. In \bibinfo{booktitle}{\emph{Proceedings of the {Second} {Workshop} on {Computing} within {Limits}}} \emph{(\bibinfo{series}{{LIMITS} '16})}. \bibinfo{publisher}{Association for Computing Machinery}, \bibinfo{address}{New York, NY, USA}, \bibinfo{pages}{1--7}.
\newblock
\showISBNx{978-1-4503-4260-5}
\urldef\tempurl%
\url{https://doi.org/10.1145/2926676.2926691}
\showDOI{\tempurl}


\bibitem[Sharma et~al\mbox{.}(2023)]%
        {sharma_2023}
\bibfield{author}{\bibinfo{person}{Vishal Sharma}, \bibinfo{person}{Neha Kumar}, {and} \bibinfo{person}{Bonnie Nardi}.} \bibinfo{year}{2023}\natexlab{}.
\newblock \showarticletitle{Post-growth {Human}–{Computer} {Interaction}}.
\newblock \bibinfo{journal}{\emph{ACM Transactions on Computer-Human Interaction}} \bibinfo{volume}{31}, \bibinfo{number}{1} (\bibinfo{date}{Nov.} \bibinfo{year}{2023}), \bibinfo{pages}{9:1--9:37}.
\newblock
\showISSN{1073-0516}
\urldef\tempurl%
\url{https://doi.org/10.1145/3624981}
\showDOI{\tempurl}


\bibitem[Sutherland(2022)]%
        {sutherland_2022}
\bibfield{author}{\bibinfo{person}{Brian Sutherland}.} \bibinfo{year}{2022}\natexlab{}.
\newblock \showarticletitle{Strategies for {Degrowth} {Computing}}. In \bibinfo{booktitle}{\emph{Eighth {Workshop} on {Computing} within {Limits} 2022}}. \bibinfo{publisher}{LIMITS}.
\newblock
\urldef\tempurl%
\url{https://www.researchgate.net/profile/Brian-Sutherland/publication/361446092_Strategies_for_Degrowth_Computing/links/62ebb1680b37cc34476d6750/Strategies-for-Degrowth-Computing.pdf}
\showURL{%
\tempurl}


\bibitem[Tainter(1988)]%
        {tainter_1988}
\bibfield{author}{\bibinfo{person}{Joseph~A. Tainter}.} \bibinfo{year}{1988}\natexlab{}.
\newblock \bibinfo{booktitle}{\emph{The collapse of complex societies}}.
\newblock \bibinfo{publisher}{Cambridge University press}, \bibinfo{address}{Cambridge New York Port Chester}.
\newblock
\showISBNx{978-0-521-38673-9}


\bibitem[Tainter(2006)]%
        {tainter_2006}
\bibfield{author}{\bibinfo{person}{Joseph~A. Tainter}.} \bibinfo{year}{2006}\natexlab{}.
\newblock \showarticletitle{Social complexity and sustainability}.
\newblock \bibinfo{journal}{\emph{Ecological Complexity}} \bibinfo{volume}{3}, \bibinfo{number}{2} (\bibinfo{date}{June} \bibinfo{year}{2006}), \bibinfo{pages}{91--103}.
\newblock
\showISSN{1476-945X}
\urldef\tempurl%
\url{https://doi.org/10.1016/j.ecocom.2005.07.004}
\showDOI{\tempurl}


\bibitem[Tomlinson and Aubert(2017)]%
        {tomlinson_2017}
\bibfield{author}{\bibinfo{person}{Bill Tomlinson} {and} \bibinfo{person}{Benoit~A. Aubert}.} \bibinfo{year}{2017}\natexlab{}.
\newblock \showarticletitle{Information {Systems} in a {Future} of {Decreased} and {Redistributed} {Global} {Growth}}. In \bibinfo{booktitle}{\emph{Proceedings of the 2017 {Workshop} on {Computing} {Within} {Limits}}} \emph{(\bibinfo{series}{{LIMITS} '17})}. \bibinfo{publisher}{Association for Computing Machinery}, \bibinfo{address}{New York, NY, USA}, \bibinfo{pages}{21--28}.
\newblock
\showISBNx{978-1-4503-4950-5}
\urldef\tempurl%
\url{https://doi.org/10.1145/3080556.3080569}
\showDOI{\tempurl}


\bibitem[Tomlinson et~al\mbox{.}(2013)]%
        {tomlinson_2013}
\bibfield{author}{\bibinfo{person}{Bill Tomlinson}, \bibinfo{person}{Eli Blevis}, \bibinfo{person}{Bonnie Nardi}, \bibinfo{person}{Donald~J. Patterson}, \bibinfo{person}{M.~Six Silberman}, {and} \bibinfo{person}{Yue Pan}.} \bibinfo{year}{2013}\natexlab{}.
\newblock \showarticletitle{Collapse informatics and practice: {Theory}, method, and design}.
\newblock \bibinfo{journal}{\emph{ACM Transactions on Computer-Human Interaction}} \bibinfo{volume}{20}, \bibinfo{number}{4} (\bibinfo{date}{Sept.} \bibinfo{year}{2013}), \bibinfo{pages}{1--26}.
\newblock
\showISSN{1073-0516, 1557-7325}
\urldef\tempurl%
\url{https://doi.org/10.1145/2493431}
\showDOI{\tempurl}


\end{thebibliography}

\end{document}